\documentclass{emulateapj}

\shorttitle{Unveiling the main heating sources in HW2}
\shortauthors{I. Jim\'enez-Serra et al.}

\usepackage{multirow}
\begin{document}

\title{Unveiling the main heating sources in the Cepheus A HW2 region} 

\author{I. Jim\'{e}nez-Serra\altaffilmark{1},
J. Mart\'{\i}n-Pintado\altaffilmark{2}, P. Caselli\altaffilmark{1}, 
S. Mart\'{\i}n\altaffilmark{3}, A. Rodr\'{\i}guez-Franco\altaffilmark{2,4}, 
C. Chandler\altaffilmark{5} and J.M. Winters\altaffilmark{6}}

\altaffiltext{1}{School of Physics \& Astronomy, E.C. Stoner Building,
The University of Leeds, Leeds, LS2 9JT, UK; I.Jimenez-Serra@leeds.ac.uk,
P.Caselli@leeds.ac.uk}

\altaffiltext{2}{Centro de Astrobiolog\'{\i}a (CSIC/INTA),
Ctra. de Torrej\'on a Ajalvir km 4,
E-28850 Torrej\'on de Ardoz, Madrid, Spain; 
jmartin.pintado@iem.cfmac.csic.es,
arturo@damir.iem.csic.es}

\altaffiltext{3}{Harvard-Smithsonian Center for Astrophysics, 
60 Garden Street, Cambridge, MA 02138; smartin@cfa.harvard.edu}

\altaffiltext{4}{Escuela Universitaria de \'Optica,  
Departamento de Matem\'atica Aplicada (Biomatem\'atica),
Universidad Complutense de Madrid,
Avda. Arcos de Jal\'on s/n, E-28037 Madrid, Spain}

\altaffiltext{5}{National Radio Astronomy Observatory,
P.O. Box O Socorro NM 87801; cchandle@nrao.edu}

\altaffiltext{6}{Institut de Radio Astronomie Millim\'etrique, 300 Rue 
de la Piscine, F-38406 St. Martin d'H\`eres, France; winters@iram.fr}

\begin{abstract}

We present high angular resolution PdBI images 
(beam of $\sim$0.33$''$) of the $J$=27$\rightarrow$26 line from several vibrational levels 
($v_7$=1 and $v_6$=1) of HC$_3$N toward Cepheus A HW2. These images 
reveal the two main heating sources in the cluster: one centered in 
the disk collimating the HW2 radio jet (the HW2 disk), 
and the other associated with a hot core 0.3$''$ northeast HW2 (the HC). 
This is the first time that vibrationally excited emission of HC$_3$N
is spatially resolved in a disk. The kinematics of this emission shows 
that the HW2 disk rotates following a Keplerian law.
We derive the temperature profiles in the two objects from 
the excitation of HC$_3$N along the HW2 disk and the HC. 
These profiles reveal that both objects are centrally heated and show 
temperature gradients. The inner and hotter regions have temperatures 
of 350$\pm$30$\,$K and 270$\pm$20$\,$K for the HW2 disk and the HC, respectively. 
In the cooler and outer regions, the temperature drops to 250$\pm$30$\,$K in 
the HW2 disk, and to 220$\pm$15$\,$K in the HC. The estimated luminosity of 
the heating source of the HW2 disk is $\sim$2.2$\times$10$^4$$\,$L$_\odot$, 
and $\sim$3000$\,$L$_\odot$ for the HC. The most massive protostar in the 
HW2 region is the powering source of the HW2 radio jet. We discuss the 
formation of multiple systems in this cluster. The proximity of the HC to 
HW2 suggest that these sources likely form a binary system of B stars,
explaining the observed precession of the HW2 radio jet.

\end{abstract}


\section{Introduction}

Cepheus A East, located at 700$\,$pc \citep{reid09} and with an IR 
luminosity of $\sim$2.5$\times$10$^{4}$$\,$L$_\odot$ \citep{evans81}, 
is a very active star forming region with signposts of 
massive star formation \citep[see e.g.][]{hug84}.
The brightest radio continuum source 
is the thermal radio jet HW2 \citep[][]{rod94}, which powers 
the northeast-southwest outflow seen in CO and HCO$^+$ 
\citep[][]{nar96,gom99}. 

Interferometric images of the molecular emission 
toward HW2 have revealed a very complex picture of the surroundings 
of the radio jet suggesting the presence of a cluster. 
\citet{mar05}, \citet{pat05}, \citet{bro07} and \citet{com07} proposed that 
the number of sources in the HW2 system could be as many as five.
Later, the higher-angular resolution observations of 
\citet{jim07} showed that the molecular gas around HW2 is 
resolved, at least, into a disk 
centered at the radio jet (the HW2 disk), and an independent hot core 
located at $\sim$0.4$''$ east HW2 (the HC). 

\citet[][]{bro07} reported vibrationally excited emission of HC$_3$N 
(hereafter, HC$_3$N$^{*}$), previously detected by \citet{mar05}, 
arising from an unresolved condensation ($\leq$1$''$; HW2-NE)
located at the same position as the HC.
The vibrational states of this molecule are excited mainly 
by IR radiation re-emitted by dust at $\lambda$$\leq$50$\,$$\mu$m.
The determination of the size of the emitting region and of the 
excitation temperature of HC$_3$N provides a good estimate of the 
luminosity of the heating object \citep[][]{devi00}. 
\citet{bro07} derived a temperature of 312$\,$K for the HW2-NE/HC source.
Assuming a size of $\sim$0.6$''$ for this object \citep[][]{mar05},
the expected IR luminosity is $\sim$2$\times$10$^4$$\,$L$_\odot$. 
Since this luminosity is similar to that measured in HW2,
and since the resolution of the \citet{bro07} images is not high
enough to discriminate between the HC and the HW2 source as the main 
heating source, the question remains whether HW2 or the HC is 
the most luminous source in the HW2 cluster.

We present high angular resolution PdBI images 
(beam of $\sim$0.33$''$) of the $J$=27$\rightarrow$26 rotational lines 
in the $v_7$=1 and $v_6$=1 vibrational 
levels of HC$_3$N, toward the Cepheus A HW2 region.
The HW2 disk and the HC are centrally heated by two massive protostars. 
The central source of the HW2 disk is the most luminous object 
in the cluster. 

\begin{figure}
\begin{center}
\includegraphics[angle=0,width=0.42\textwidth]{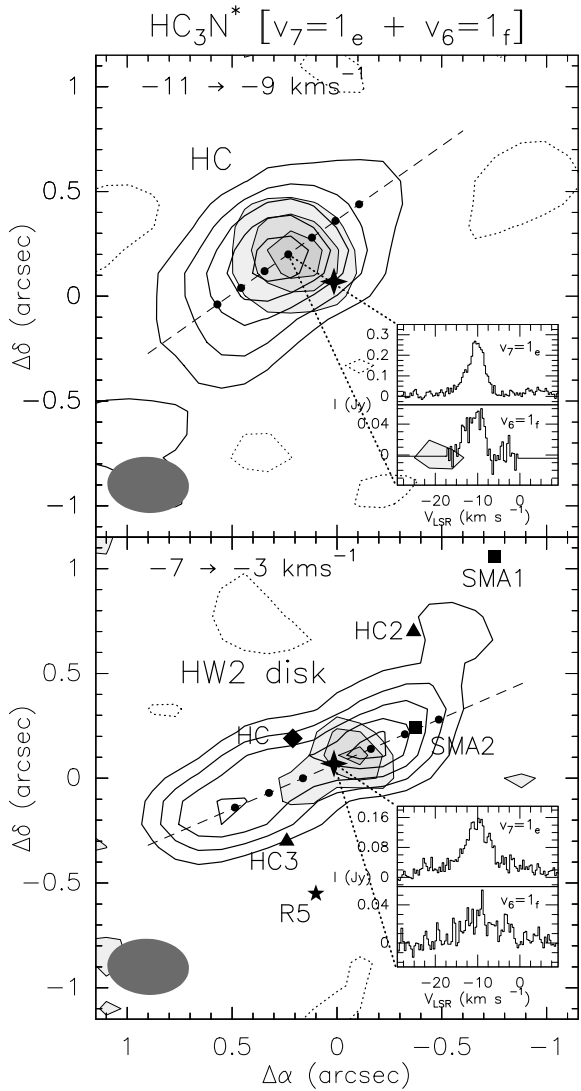}
\caption{Integrated intensity images of the HC$_3$N$^*$ 
$J$=27$\rightarrow$26 $v_7$=1$_e$
(contours) and $v_6$=1$_f$ (grey scale and thin contours) line emission
observed toward the HC ($-$11 to $-$9$\,$km$\,$s$^{-1}$; 
upper panel) and the HW2 disk (from $-$7 to $-$3$\,$km$\,$s$^{-1}$; lower panel). 
For the HC, the first contour and step level are 30 (3$\sigma$) 
and 50$\,$mJy$\,$km$\,$s$^{-1}$ for the $v_7$=1$_e$ map, and 10 (2$\sigma$) and 
10$\,$mJy$\,$km$\,$s$^{-1}$ for the $v_6$=1$_f$ image. For the HW2 disk,
the first contour and step level are 12 (2$\sigma$) and 12$\,$mJy$\,$km$\,$s$^{-1}$
for the $v_7$=1$_e$ emission, and 12 (2$\sigma$) and 6$\,$mJy$\,$km$\,$s$^{-1}$ 
for the $v_6$=1$_f$ line. Negative contours correspond to the 3$\sigma$ level. 
The central coordinates are [$\alpha(J2000)$=22$^{h}$56$^{m}$17.98$^s$ 
and $\delta(J2000)$=+62$^{\circ}$01$'$49.5$''$]. Filled cross shows the 
expected location of HW2 \citep{cur06}.
The dashed lines and filled circles 
indicate the direction and positions at which the
HC$_3$N$^*$ excitation is calculated (Sec.$\,$$\ref{phys}$). Beam size 
is shown at lower left corner. The spectra of the $v_7$=1$_e$ and $v_6$=1$_f$ lines
toward the positions of the HC and the HW2 source are reported in the inner panels at lower 
right corner. We also show the location of the HC (filled 
diamond), SMA1 and SMA2 \citep[filled squares;][]{bro07},
HC2 and HC3 \citep[filled triangles;][]{com07}, and R5
\citep[filled star;][]{torr01} in the lower panel of this figure.}
\label{f1}
\end{center}
\end{figure}

\section{Observations and Results}
\label{res}

The $J$$\,$=$\,$27$\rightarrow$26 $v_7$=1$_e$ and 1$_f$ lines 
($E_u/k$=487$\,$K), and the $v_6$=1$_e$ and 1$_f$ transitions 
($E_u/k$=883$\,$K) \footnote{The $v_7$ and $v_6$ vibrational 
levels correspond to the bending modes of the C-C$\equiv$N and C$\equiv$C-C 
bonds.} 
were observed simultaneously with the PdBI in the A configuration\footnote{Based 
on observations carried out with the IRAM Plateau de 
Bure Interferometer. IRAM is supported by INSU/CNRS (France), MPG (Germany) 
and IGN (Spain)}. 
The correlator setup provided spectral resolutions of $\sim$80 and 160$\,$kHz,
i.e.$\,$$\sim$0.1-0.2$\,$km$\,$s$^{-1}$ at 246$\,$GHz. The 
synthesized beam size was 0.39$''$$\times$0.28$''$ 
with a position angle (P.A.) of 85$^{\circ}$. We used 3C454.3 
(17$\,$Jy) and 3C273 (11$\,$Jy) as bandpass calibrators; 
MWC349 (1.3$\,$Jy), as flux density calibrator; 
and 1928+738 (0.8$\,$Jy) and 0212+735 (0.7$\,$Jy), as phase calibrators. 
Calibration, continuum 
subtraction, imaging and cleaning were done with the GILDAS 
package\footnote{See http://www.iram.fr/IRAMFR/GILDAS.}. 

The four HC$_3$N$^*$ $v_7$=1 and $v_6$=1 transitions were detected toward HW2. 
In Figure$\,$\ref{f1}, we show the integrated intensity maps of the 
$v_7$=1$_e$ (contours) and $v_6$=1$_f$ 
(grey scale and thin contours) emission from   
$-$11 to $-$9$\,$km$\,$s$^{-1}$ (the HC; upper panel), and from 
$-$7 to $-$3$\,$km$\,$s$^{-1}$ (the HW2 disk; lower panel). 
These lines are not blended, and therefore were mainly used in the analysis
of our data. Figure$\,$\ref{f1} also shows the $v_7$=1$_e$ and $v_6$=1$_f$ 
spectra observed toward the HC and the HW2 source.
In agreement with the SO$_2$ images of \citet{jim07}, 
the HC$_3$N$^*$ emission is resolved into two main
molecular condensations: one centered at the HC, and the other 
associated with the protostellar disk around the HW2 jet, the HW2 disk.
For completeness, we show in Figure$\,$\ref{f1} (lower panel) 
the location of sources SMA1, SMA2 \citep[filled squares;][]{bro07}, 
HC2, HC3 \citep[filled triangles;][]{com07}, and R5 
\citep[filled star;][]{torr01} also reported in the region.

\subsection{The Hot Core (HC)}
\label{hc}

The HC$_3$N$^*$ emission from $-$11
to $-$9$\,$km$\,$s$^{-1}$ (upper panel, Figure$\,$\ref{f1}) 
is relatively compact and mainly arises from the HC.  
The central radial velocity of this condensation is
$v_{LSR}$=$-$10$\,$km$\,$s$^{-1}$, and its peak emission is located
at $\sim$0.3$''$ north-east HW2. 
This position is similar to that reported 
by \citet{mar05} for the HC, and consistent with the location of the HW2-NE 
source \citep{bro07}. The deconvolved size of the $v_7$=1$_e$ line emission
is 0.4$''$$\times$0.7$''$ (270$\,$AU$\,$$\times$$\,$480$\,$AU), and for 
the $v_6$=1$_f$ line, 0.25$''$$\times$0.4$''$ 
(170$\,$AU$\,$$\times$$\,$270$\,$AU).
The HC is likely the powering source of the small-scale SiO outflow 
reported by \citet{com07} toward HW2.

\subsection{The HW2 disk}
\label{disk}

From $-7$ to $-3$ km$\,$s$^{-1}$, the HC$_3$N$^*$ 
emission shows an elongated structure centered on HW2 that resembles the 
SO$_2$ disk reported by \citet{jim07}. To our knowledge, this is the 
first time that high-excitation HC$_3$N$^*$ lines are spatially resolved 
toward a protostellar disk. The HW2 disk is centered at 
$v_{LSR}$=$-$5$\,$km$\,$s$^{-1}$, which gives a velocity difference 
between the HC and this object of $\sim$5$\,$km$\,$s$^{-1}$. 
This difference has also been observed in the circumstellar 
molecular gas around HW2 at larger scales 
\citep[][]{cod06}. 

While the HC$_3$N$^*$ $v_7$=1$_e$ emission arises from all 
along the disk (deconvolved size of 1.4$''$$\times$0.18$''$, 
950$\,$AU$\times$$\,$120$\,$AU), the $v_6$=1$_f$ line is restricted to 
the inner and hotter regions closer to the HW2 source (size of 
$\leq$0.18$''$$\,$$\times$$\,$0.4$''$, $\leq$120$\,$$\times$$\,$270$\,$AU).
The orientation of the $v_7$=1$_e$ and $v_6$=1$_f$ line emission 
(P.A.$\simeq$ 115$^{\circ}$) is roughly perpendicular to the HW2 jet 
\citep[P.A.$\simeq$ 46$^\circ$;][]{rod94}. This orientation, although not 
exactly the same, is similar to that seen in SO$_2$, CH$_3$CN and NH$_3$ for 
the same velocity range \citep{torr07}. We note that \citet{bro07} 
reported larger differences in the disk orientation for different 
molecular tracers. However, their molecular emission 
images were obtained for a larger velocity range 
from $\sim$$-$13 to 0$\,$km$\,$s$^{-1}$. In any case, the
discrepancies in the disk orientation could be produced by either chemical or 
excitation effects.

\begin{figure}
\includegraphics[angle=270,width=0.47\textwidth]{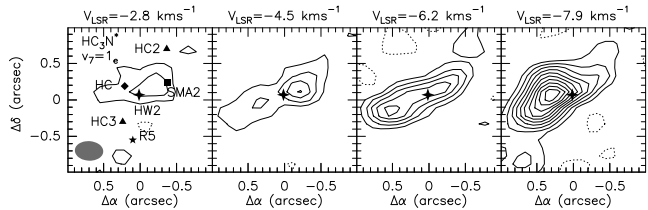}
\caption{Images of the 
HC$_3$N$^*$ $v_7$=1$_e$ emission 
observed toward HW2 for the velocity intervals from 
$-$2 to $-$3.7 km$\,$s$^{-1}$, $-$3.7 to $-$5.4 km$\,$s$^{-1}$,  
$-$5.4 to $-$7.1 km$\,$s$^{-1}$, and $-$7.0 to $-$8.9 km$\,$s$^{-1}$. 
The central radial velocities of these intervals
are shown in the upper part of the panels.  
The first contour and step levels 
are 16 (2$\sigma$) and 16$\,$mJy$\,$km$\,$s$^{-1}$. 
Symbols are as in Figure$\,$\ref{f1}. Beam size is shown in the lower 
left corner.} 
\label{f2}
\end{figure}

The integrated intensity maps of the $v_7$=1$_e$ line for the 
velocity intervals of \citet{jim07}, are shown in Figure$\,$\ref{f2}. 
The kinematics of this emission are similar to those observed from 
SO$_2$ for the HW2 disk. The red-shifted 
HC$_3$N$^*$ emission is brighter toward the northwest of HW2, and 
progressively moves to the southeast for blue-shifted 
velocities. The HC$_3$N$^*$ emission at 
$V_{LSR}$=$-$7.9 km$\,$s$^{-1}$ is clearly dominated by the HC.  
The velocity difference between the north-west and the south-east part of
the HW2 disk is $\sim$5$\,$km$\,$s$^{-1}$ over a projected distance of 
$\sim$1000$\,$AU. 

\begin{figure}
\includegraphics[angle=270,width=0.47\textwidth]{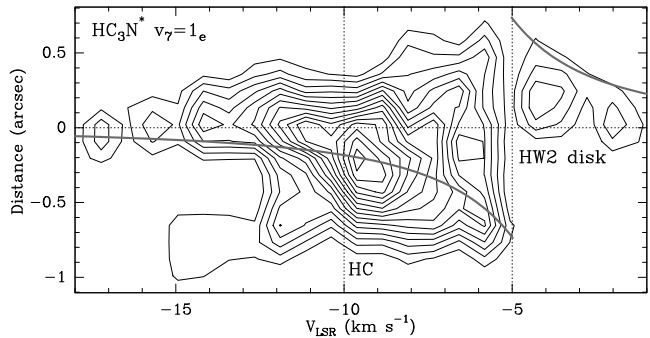}
\caption{P-V diagram of the HC$_3$N$^*$ $v_7$=1$_e$ emission along the 
HW2 disk. First contour and step level are 15 (3$\sigma$) and 10$\,$mJy, respectively.
Distances are calculated with respect to the HW2 source 
\citep[horizontal dotted line;][]{cur06}. Vertical dotted lines show the 
central radial velocities of the HW2 disk ($-$5$\,$km$\,$s$^{-1}$) and of 
the HC ($-$5$\,$km$\,$s$^{-1}$). The Keplerian rotation velocity for 
a central source with mass 18$\,$M$_\odot$, a disk size of $\sim$1000$\,$AU, 
and an inclination angle of 62$^\circ$, is also shown (thick grey line).}
\label{f3}
\end{figure}

In Figure$\,$\ref{f3}, we show the P-V diagram of the $v_7$=1$_e$ emission 
along the HW2 disk, after smoothing the data to a velocity resolution 
of 0.76$\,$km$\,$s$^{-1}$. We also superpose the Keplerian rotation 
velocity curve for a star of 18$\,$M$_\odot$, a disk size of $\sim$1000$\,$AU, 
and an inclination angle of the disk axis with respect to the line of sight 
of $\sim$62$^\circ$ \citep{pat05}. Although the morphology of the 
HC$_3$N$^*$ emission is very complex, the agreement of the data 
with Keplerian rotation (thick grey line) is particularly 
good for the red-shifted part of the HW2 disk. The blue-shifted part
is contaminated by emission from the HC and shocked gas 
\citep[][]{jim07}. 
Therefore, the HW2 disk seems to rotate following a Keplerian law. 
The central mass of 18$\,$M$_\odot$ is consistent with that estimated 
from the source size and the HC$_3$N excitation temperature for the HW2 disk 
(Section$\,$\ref{exc}).

In contrast with the HW2 disk scenario, \citet{bro07} reported chemical 
differences within the molecular structure around HW2 at 
$\sim$1$''$-2$''$ scales. These authors proposed that this 
feature could be instead due to the superposition of two
independent objects in the plane of the sky. Although this possibility cannot 
be ruled out, there are several observational evidences that 
favour the disk scenario: 
i) HC$_3$N$^*$ (this work), SO$_2$ and NH$_3$  
\citep{jim07,torr07}, show a continuous structure whose kinematics 
are consistent with Keplerian rotation; ii) the observed 
HC$_3$N$^*$ peaks do not seem to coincide with SMA1, 
HC2, HC3 or R5; for SMA2, this source falls $\sim$0.25$''$ northwest the 
red-shifted HC$_3$N$^*$ peak (we note that the uncertainty in absolute 
position is of $\leq$0.1$''$); iii) the dust continuum emission 
delineates a similar structure to that seen in HC$_3$N$^*$, SO$_2$ or
NH$_3$ \citep{torr07}; and iv) the 
spatial distribution of the H$_2$O and CH$_3$OH masers 
\citep{torr96,sugi07}, is consistent with the presence of a disk with 
radius $\sim$600-700$\,$AU. Therefore, we propose that the observed 
chemical differences within the HW2 disk could be produced by  
different external physical conditions, or by the presence of 
other hot core sources in the vicinity of HW2.

\section{Temperature profiles and HC$_3$N column densities}
\label{phys}

We can combine the $v_6$=1 and the $v_7$=1 lines of HC$_3$N to derive the
excitation temperature of the hot gas along the HC and the HW2 disk 
(Figure$\,$\ref{f1}) by means of Boltzmann diagrams. These diagrams assume 
LTE and optically thin emission. In case the $v_7$ lines were 
moderately optically thick, the derived excitation temperatures should be 
considered as upper limits.

\begin{figure}
\includegraphics[angle=270,width=0.47\textwidth]{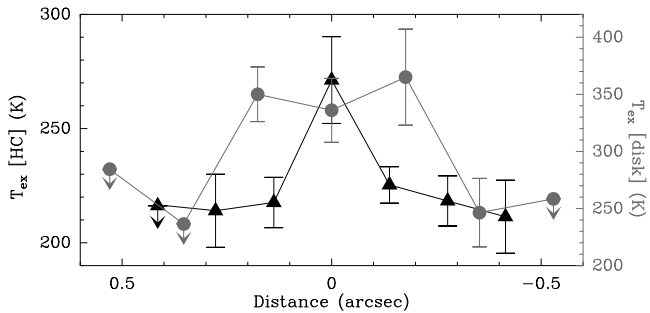}
\caption{Radial distribution of the HC$_3$N excitation temperatures, 
with their errors, across the HC (filled triangles) and the HW2 disk (filled circles). 
The temperature scale for the HC is shown on the left, and for the 
HW2 disk, on the right. Arrows indicate 
the upper limits to the excitation temperatures of HC$_3$N.}
\label{f4}
\end{figure}

Figure$\,$\ref{f4} shows the radial profile of the excitation 
temperatures of HC$_3$N toward the HC (filled triangles)
and the HW2 disk (filled circles). The errors associated with 
these temperatures are $\leq$12\%. For the HC, the temperature profile 
is centrally peaked with a maximum value of 
270$\pm$20$\,$K, and drops to excitation temperatures of 
$\leq$225$\,$K within the inner 0.15$''$. This is consistent with the 
idea that the HC is internally heated by an IR source 
\citep[see Figure$\,$3 of][]{devi02}, and suggests that the hot gas is highly
concentrated around the protostar. 
The temperature of 270$\,$K is larger than that derived by 
\citet[][$\sim$160$\,$K]{mar05} from single-dish HC$_3$N data, but  
lower than that calculated by 
\citet[][$\sim$312$\,$K]{bro07}. In the former, 
dilution effects could account for the temperature 
discrepancies. In the latter, 
the excitation temperature of $\sim$312$\,$K is only an upper limit,
since the low lying HC$_3$N$^*$ lines are likely optically thick 
\citep{bro07}.

From the $v_7$=1$_e$ integrated line flux ($\sim$0.7$\,$Jy$\,$km$\,$s$^{-1}$), 
and assuming a temperature of $\sim$220$\,$K, the
derived HC$_3$N column density in the HC is of $\sim$10$^{16}$$\,$cm$^{-2}$. 
The partition function of the HC$_3$N $v_7$ level has been calculated from 
the one in the ground vibrational state, but multiplied by $\sim$0.3. 
This corresponds to the factor $exp[-(290\,K/220\,K)]$, where 290$\,$K
is the energy of the fundamental rotational level in the $v_7$ state and 
220$\,$K is the derived excitation temperature.
If we consider an HC$_3$N abundance of $\sim$10$^{-8}$ \citep[as for 
the Orion hot core;][]{devi02}, the estimated H$_2$ column density is 
$\sim$10$^{24}$$\,$cm$^{-2}$. Assuming a size of $\sim$0.5$''$, this leads 
to a circumstellar mass of $\sim$0.6$\,$M$_\odot$ for the HC, which 
is consistent with that derived by \citet{mar05} from SO$_2$. 

For the HW2 disk, the maximum excitation of HC$_3$N is found at 
$\sim$0.2$''$ northwest and southeast of the radio jet
(350$\pm$40 and 365$\pm$25$\,$K, respectively), suggesting that the 
hot gas is distributed in an inner disk of radius 
$\sim$0.2$''$ (140$\,$AU; as estimated from Figure$\,$\ref{f4}).
Outside this disk, the temperature 
falls by more than 100$\,$K. From the 
$v_7$=1$_e$ integrated line flux ($\sim$0.4$\,$Jy$\,$km$\,$s$^{-1}$),
and assuming a temperature of $\sim$250$\,$K, 
the derived column density of HC$_3$N is
$\sim$3$\times$10$^{16}$$\,$cm$^{-2}$.
By using Eq.$\,$2 of \citet{jim07},
and assuming an HC$_3$N abundance\footnote{HC$_3$N is likely 
being photo-dissociated in the disk. The
HC$_3$N abundance should be a factor of 10-30 lower 
than that found in the HC \citep[e.g.][]{rofr92}.} of 
$\sim$3$\times$10$^{-10}$-10$^{-9}$, 
the estimated mass for the HW2 disk is $\sim$0.4-1$\,$M$_\odot$. This mass 
is similar to that obtained by \citet{pat05}, \citet{jim07} and \citet{torr07}.

\section{On the heating of the HC and the HW2 disk}
\label{exc}

In the case of radiative excitation, 
the HC$_3$N gas temperatures are a good estimate of the 
dust temperatures because the continuum emission of the HC 
and the HW2 disk at 20 and 45$\,$$\mu$m is likely optically thick. 
If we assume an H$_2$ column density of $\sim$10$^{24}$$\,$cm$^{-2}$, 
a gas-to-dust mass ratio of 100, and dust opacities of 3360 and 1810$\,$cm$^2$g$^{-1}$
at 20 and 45$\,$$\mu$m for both objects \citep{oss94}, the derived 
optical depths are $\geq$60. This explains the large obscuration seen
toward HW2 at 24.5$\,$$\mu$m by \citet{dewit09}. 
Since gas and dust are thermally coupled,
we can then estimate the IR luminosity of the central source by 
using the Stefan-Boltzmann law \citep[][]{devi00}.
From the $v_7$=1 emission size ($\sim$0.5$''$), and
considering spherical symmetry for the HC, we derive an IR luminosity of 
$\sim$3000$\,$L$_\odot$ for a dust temperature of 220$\,$K. 
This luminosity, which is consistent with that obtained by \citet{mar05},
would correspond to a ZAMS B2-type star 
\citep[$\sim$10$\,$M$_\odot$;][]{pan73}.   

If we now consider an edge-on disk geometry with radius 475$\,$AU and 
height 120$\,$AU for the HW2 disk, we estimate an IR luminosity of 
2.2$\times$10$^4$$\,$L$_\odot$ for a dust temperature of 250$\,$K. This 
is consistent with a ZAMS B0 star of $\sim$18$\,$M$_\odot$ for the HW2  
source \citep[][]{pan73}. 
   
\section{Discussion}
\label{dis}

The high-angular resolution HC$_3$N$^*$ images 
toward Cepheus A HW2 have shown that this multiple system contains 
two massive protostars. 
The powering source of the HW2 disk stands out among them with 
an IR luminosity of 2.2$\times$10$^4$$\,$L$_\odot$. 
The number of ionizing photons for a B0 star, as inferred from the heating 
\citep[$\sim$10$^{47}$$\,$s$^{-1}$;][]{pan73}, is similar 
to that estimated by \citet{hug95} from the radio continuum emission of the 
HW2 radio jet ($\gg$5$\times$10$^{46}$$\,$s$^{-1}$). This indicates that the 
HW2 source is the most massive protostar in the region.

The mass of the HW2 disk, as derived from HC$_3$N,
is relatively high ($\sim$0.4-1$\,$M$_\odot$) suggesting that this object is at 
an early stage of evolution. The dynamical age of the 
CO/HCO$^+$ outflow \citep[$\sim$5$\times$10$^3$$\,$yr;][]{nar96} is
short compared to the lifetime of a photo-evaporating disk for a 
$\sim$18$\,$M$_\odot$-star \citep[$\leq$10$^5$$\,$yr;][]{gor09},   
suggesting that the HW2 disk probably constitutes the most massive 
disk associated with B stars detected so far \citep{fue03}. 
The chemical differences observed within the HW2 disk \citep{bro07},
could be due to different external physical conditions or to 
other hot core sources present within this multiple system.

The derived temperature profile for the HC indicates that this object 
is centrally heated by a protostar with an IR luminosity of 
$\sim$3000$\,$L$_\odot$.
External heating of the HC would require luminosities of 
$\geq$10$^5$$\,$L$_\odot$ for the HW2 source as pointed out by 
\citet{mar05}. Shock heating could not account for the luminosity of 
the HC since the mechanical luminosity of the outflows in the region
is only $\sim$40$\,$L$_\odot$ \citep[][]{nar96}. Therefore, the HC hosts
a massive protostar. Since this object has not yet ionized its 
surroundings, the HC is at an earlier stage of evolution than the 
HW2 source. 

Massive stars are usually found in binary systems \citep{mas98}. 
The proximity of the HC to HW2 ($\sim$200$\,$AU) could be interpreted as
a binary system. Coeval formation is likely the main mechanism for the 
formation of low-mass T-Tauri binaries \citep{ghe97}. 
However, the orbits of these stars are preferentially aligned with the 
circumstellar disks of the primary stars \citep{jen04}, 
which contrasts with the non-coplanarity of the HC/HW2 disk system. 
Alternatively, the HC could have formed after 
the fragmentation of the HW2 disk \citep[][]{krum09}, but this would
require disk-to-total-stellar mass ratios of $\sim$0.1-0.2, i.e. 
factors of 3 and 6 larger than that observed in HW2 ($\sim$0.03). 

As shown by \citet{cun09}, the capture of a massive companion (the HC) 
by the HW2 disk/protostar system in an eccentric orbit
could explain the observed precession of the HW2 jet. 
If we assume an averaged orbital period of 
$\sim$3700$\,$yrs for the binary \citep[Figure$\,$7 in][]{cun09}, 
the expected mass enclosed within the system would be  
$\sim$25$\,$M$_\odot$ for a velocity difference of 
5$\,$km$\,$s$^{-1}$ and an inclination angle of 62$^\circ$. This mass is 
similar to the sum of the masses of the HC and the HW2 source as 
derived from HC$_3$N$^*$.


Finally, we cannot rule out the idea that the formation of HW2 
has triggered the formation of the HC and other objects in the region. 
Cepheus A HW2 therefore would resemble the case of
SgrB2, where previous episodes of star formation 
triggered the formation of new clusters of massive 
hot cores \citep{devi00}.  

In summary, we report the detection of HC$_3$N$^*$ emission
toward the HC and the HW2 disk in Cepheus A HW2. 
The HC$_3$N$^*$ images show that these objects are centrally heated
by massive protostars (of 18 and 10$\,$M$_\odot$ for the HW2 source and 
the HC, respectively). 
Since they appear to be very close (200$\,$AU), we propose that these 
objects form a binary system of massive stars. 

\acknowledgments

We acknowledge the IRAM staff for the support provided during the observations. 
We thank two referees for their useful comments that helped 
to improve the paper, and W. J. de Wit and E. R. Parkin for fruitful 
discussions. JMP acknowledges the Spanish MEC for the support provided 
through projects number ESP2004-00665, ESP2007-65812-C02-01 and
``Comunidad de Madrid'' Government under PRICIT project
S-0505$/$ESP-0277 (ASTROCAM).

\end{document}